\documentclass[twoside,agupp]{aguplus}
\usepackage{psfig,graphics,mjh,graphpap,epic,eepic,color,multirow}
\usepackage{lineno}

\figmarkoff

\lefthead{HEAVNER ET AL.}
\righthead{NUV/BLUE SPRITE SPECTRAL OBSERVATIONS}
\articleid{1}{13}
\paperid{}
\ccc{}
\cpright{AGU}{2010}
\authoraddr{M.J. Heavner, Los Alamos National Lab, MS D-436, Los Alamos NM 87544
(email: heavner@lanl.gov).}
\authoraddr{D.D. Sentman, D.R. Moudry, and E.M. Wescott
GI-UAF, Fairbanks, AK 99775 (email: [dsentman,drm,gene.wescott]@gi.alaska.edu).}
\authoraddr{J.S. Morrill amd C.L. Siefring, Naval Research Laboratory,
Washington DC 20375 (email: [morrill@shogun,siefring@ccs].nrl.navy.mil).}
\authoraddr{E.J. Bucsela, SRI International, 333 Ravenswood Ave, 404-57, Menlo Park, CA 94025-3493
  (email: eric.bucsela@sri.com).}

\slugcomment{To appear in  {\it Journal of Geophysical Research}, 2010.}

\begin{document}
\title{NUV/Blue spectral observations of sprites in the 320-460 nm
region: ${\mathrm N_2}$ (2PG) Emissions}
\author{M.J. Heavner}
\affil{Los Alamos National Lab, Los Alamos NM 87544}
\author{J.S. Morrill and C. Siefring}
\affil{Naval Research Laboratory, Washington DC 20375}
\author{D.D. Sentman, D.R. Moudry, and E.M. Wescott}
\affil{Geophysical Institute, University of Alaska Fairbanks,
  Fairbanks AK 99775}
\author{E.J. Bucsela}
\affil{SRI International, Menlo Park, CA 94025}

\begin{abstract}
A near-ultraviolet (NUV) spectrograph (320-460~nm) was flown on the EXL98
aircraft sprite observation campaign during July~1998.  
In this wavelength range video rate (60 fields/sec) spectrographic
observations found the NUV/blue emissions to be predominantly \nttp.
The negligible level of \nton\ present in the spectrum is confirmed by
observations of a co-aligned, narrowly filtered 427.8~nm imager and is
in agreement with previous ground-based filtered photometer
observations.  The synthetic spectral fit to the observations
indicates a characteristic energy of $\sim$1.8~eV, in agreement with our
other NUV observations.
\end{abstract}

\begin{article}
\section{Introduction}
Sprites are brief ($\sim$10~ms) optical phenomena that occur above
thunderstorms at altitudes between 40-95~km, which have been
well documented only durring the last two decades \citep[see][]{sentman98, sentman2008:spriteplasma}.
Broadband, spectral, and filtered photometric observations have been
made across the visible wavelengths.  Color camera observations show
that sprites are primarily red, becoming blue at the lower altitudes
\citep{sentman95}.

Ground-based spectroscopic measurements of sprites in the optical
bandpass of 400-850 nm \citep{mende95,hampton96} show the strongest
optical emission from sprites is \ntop, first positive molecular
nitrogen emission, which is excited via electron impact.  Analysis of
red spectral observations estimated a 1~eV Boltzmann electron
distribution would produce the observed spectral vibrational
distribution \citet[]{green96}.  \citet{milikh97} independently modeled
the observed red spectra with similar results.  
Based on modelling of the red spectral observations of \ntop
\ emissions, several early predictions of blue emissions in sprites
were published \citep{pasko97, cho98, morrill98, dupre98}.  More
recent modeling efforts have eximated other \nt \ emissions such as the
\nt LBH in the UV \citep{liu2005:LBH}.  These predictions are compared
with the reported observations in the discussion section below.

Analysis of one red spectral observation of a sprite required \ntm \
(ionized molecular nitrogen Meinel) emissions for a best fit to the
observations \citep{morrill98}.  The \ntop \ and \ntm \ observations
and analysis are the subject of another paper \citep{bucsela99prep}.
Photometric studies of sprites from the ground by \citet{armstrong98}
and \citet{suszcynsky98} found the blue neutral molecular nitrogen
second positive (\nttp) and \nton\ are generally of shorter duration
($\sim$5~ms) than the \ntop\ red emissions (which may last up to
$\sim$100~ms).  Furthermore, according to the photometer data, the
ionized \nton\ are more brief ($\sim$1~ms) than the \nttp\ emissions.
In addition to the poor atmospheric transmission in the blue, due to
Rayleigh scattering (see Fig.~1 of \citet{morrill98}), the shorter
duration of the blue emissions compared to the red emissions is a
factor making blue observations of sprites more difficult.  The
shorter duration of the blue emissions is simply a factor of the
shorter lifetime (nanoseconds rather than microseconds, as shown in
Table~1) of the \ntc and \ntpb upper states of the \nttp and \nton
blue bands.  Other recent analysis of space based observations by
\cite{Kuo:2005p2636} has examined 2PG/1NG emission ratios and yield
characteristic energies between 4.5 and 6.5~eV.

\begin{figure}[t]
\begin{picture}(240,260)
\put(15,15){
  \resizebox{220pt}{240pt}{\includegraphics*{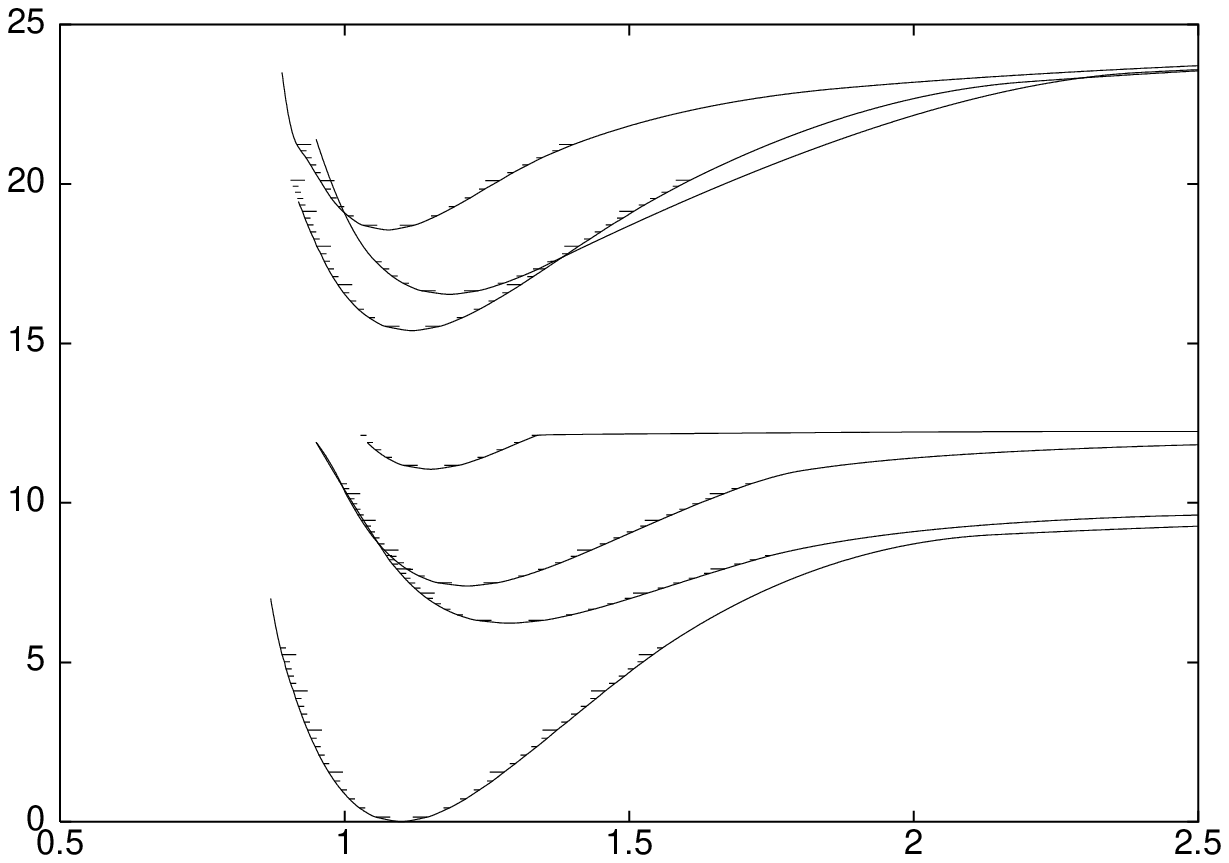}}}
\put(45,110){\Large \nt} \put(175,80){\large \ntnx}
\put(170,117){\large \ntna} \put(113,117){\large \ntnb}
\put(104,140){\large \ntnc} \put(45,182){\Large \ntp}
\put(120,170){\large \ntpnx} \put(150,194){\large \ntpna}
\put(100,225){\large \ntpnb} \put(60,0){\large Internuclear Distance
  (\AA)} \put(0,90){\rotatebox{90}{\large Potential Energy (eV)}}
\end{picture}
\caption{Grotrian Energy Diagram for \nt, with selected
electronic states plotted.  The \ntop, \nttp, \nton, and \ntm \
transitions are indicated.  This figure
was made using Tables 80 and 81 from \citet{lofthus77}.}
\label{nitro:grotrian}
\end{figure}

The primary band systems of molecular nitrogen which lead to
near-optical emissions (300-1000~nm) are the \ntop\ and \nttp\ neutral
and the \nton\ and \ntme\ ionized molecular nitrogen emissions.
Table~1 provides an overview of nitrogen and oxygen optical emissions
observed in the aurora.  The transitions associated with the nitrogen
band systems are the $\ntb\rightarrow\nta$ transition yielding the
\ntop\ emission, the $\ntc\rightarrow\ntb$ transition yielding the
\nttp\ emission, the $\ntpb\rightarrow\ntpx$ transition giving the
\nton\ emission, and the $\ntpa\rightarrow\ntpx$ transition giving the
\ntme\ emission.  The potential energy curves of both neutral and
ionized molecular nitrogen electronic states involved in the above
transitions and the ground neutral state are also shown in
Fig.~\ref{nitro:grotrian}.  The \nttp\ and \nton\ bands occur at NUV
and blue wavelengths ($<450$ nm, with brightest emissions $<400$ nm),
while the \ntop\ and \ntme\ bands occur at red and near infrared (NIR)
wavelengths.

Because of the uncertainties associated with \ntme\ quenching and
energy transfer, characterization of \ntp\ (ionized molecular
nitrogen) is best accomplished by observation of \nton\ at shorter
wavelengths.  One reason for investigating the \nttp\ emission rather
than the dominant red \ntop\ emission is that the upper state of
\ntop, \ntb, is partially filled via cascading from the \ntc\ state
(this cascading is evidence of the \nttp\ emissions).  Also the \ntop\
upper state is more strongly affected by collisional processes than
\nttp\ \citep{morrill96} complicating the accurate analysis of \ntop.
The problems of both cascading and quenching are less severe issues
for the \ntpb\ and \ntc\ upper electronic states \citep{morrill96}.
Additionally, \ntop\ has the lowest excitation energy threshold of the
four band systems discussed, so emissions from other groups are
evidence of higher energy processes.  Specifically, from Table~1, the
\ntop\ threshold energy is 7.5~eV comapred to 11.18~eV for \nttp\,
18.56~eV for \nton\, and 16.54~eV for \ntm.

In this paper we present an analysis of the recent blue
spectral observations made during the EXL98 aircraft sprite-observing
mission over the U.S. Midwest.  Other papers will present further
analysis associated with both specific ionization issues and the
\nttp\ to \nton\ ratio \citep{bucsela99prep, morrill2002:e_energy}.
The EXL98 filtered blue imaging of sprites associated with ionized
\ntp\ emissions are presented elsewhere \citep{heavnerth}.

\section{Instrumentation}

The \underline{E}nergetics of Upper Atmospheric E\underline{x}citation
by \underline{L}ightning (EXL98) campaign was a collaboration between
the University of Alaska, Fairbanks (UAF), the Naval Research
Laboratory (NRL), and the Air Force Research Laboratory.  This program
was designed to investigate the energy budget of middle and upper
atmospheric processes excited by lightning using several imaging and
photometric observations from a Gulfstream II aircraft.  In this paper
we present results from a `427.8~nm' filtered imager and a bore
mounted unfiltered imager both have a 9.3$\deg$ x 7.0$\deg$
field-of-view (FOV) as well as a NUV/Blue spectrograph.

Two imaging instruments were designed specifically to observe neutral
\nttp\ and ionized \nton\ emissions.  A Dage/MTI SIT VE1000 camera
with a Fuji f/0.7 lens was used to measure images of the \nton(0,1)
band emissions with a narrow passband filter at approxiately 427.8~nm.
The `427.8'~nm filter is a carefully selected filter for observation
of \nton.  It is a 1.44~nm FWHM filter actually centered at 428.3 nm,
specifically selected slightly red of the \nttp\ 426.8~nm emissions
which would contaminate the filter if it had been centered on
427.8~nm.

A GEN IIUV V53-1845 Video Scope camera and a Lyman Alpha II f/1.7 lens
system was was optimized for NUV observations.  Originally the camera
was used as a filtered imager, with a 340.7~nm filter, in order to
make observations of \nttp\ (neutral) emissions.  During the later
portion of the campaign, this instrument was refitted as a NUV
spectrograph.  The NUV spectrograph consisted the GEN IIUV V5301845
Video Scope camera as the detector for light which passed through a
Jobin Yvon-Spex CP200 f/2.9 spectrograph and a 133 line/mm grating.
The primary input lens focused the incoming light on an 8~mm
collimating entrance slit.  The slit was oriented horizontally in
order to increase the probability of sprite observations.  The data
was recorded at 60 fields per second and all cameras on aircraft and
ground were GPS synchronized providing video images with a
common time base to 0.5 $\mu$s or better.

Other instrumentation and the overall EXL98 campaign are described by
\citet{siefring:2009NIR}.

\section{Observation}

\begin{figure}[t]
  \begin{picture}(230,350)
    \put(0,0){\resizebox{230pt}{!}{\includegraphics*{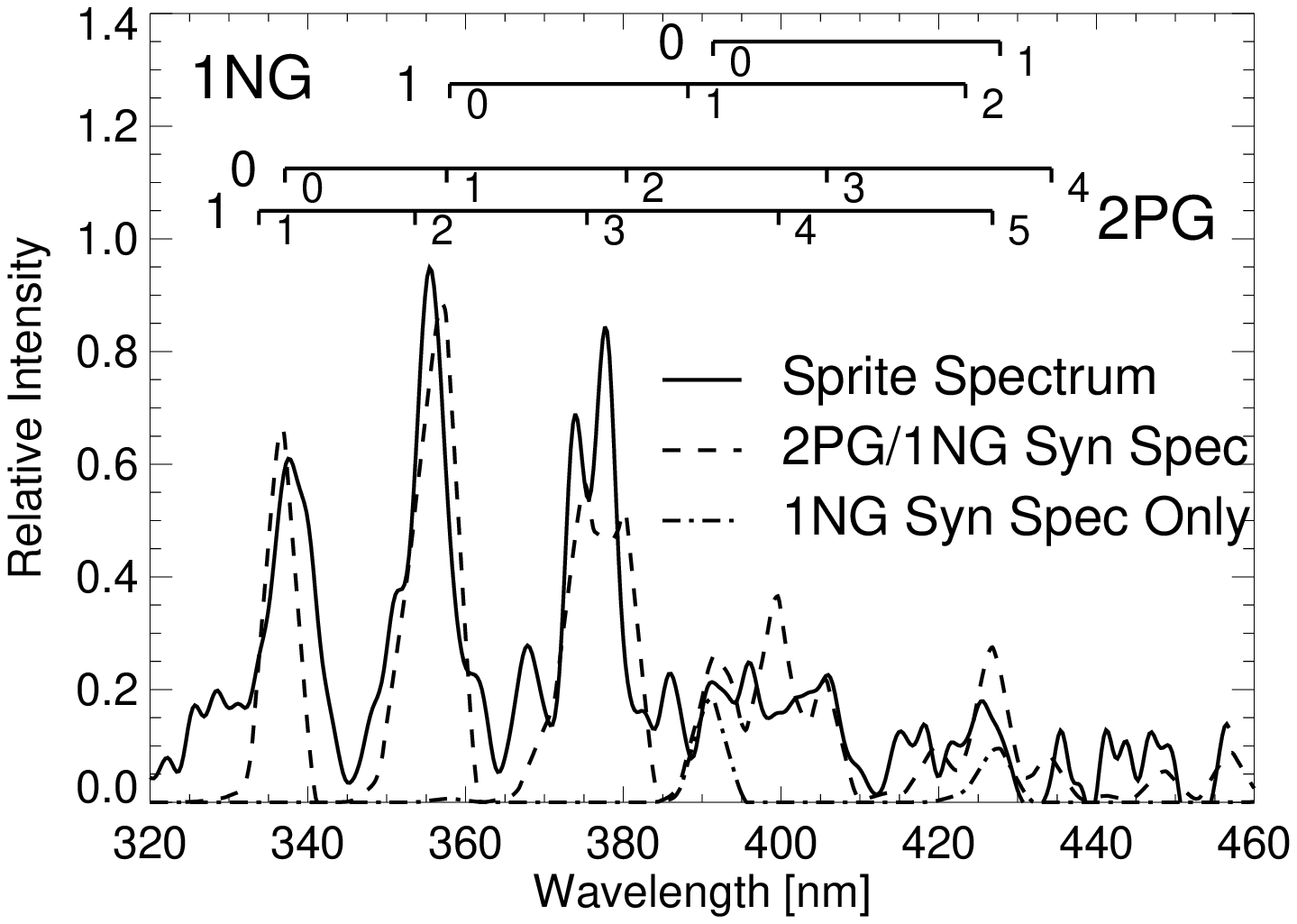}}}
    \put(10,180){\resizebox{100pt}{!}{
        \includegraphics*{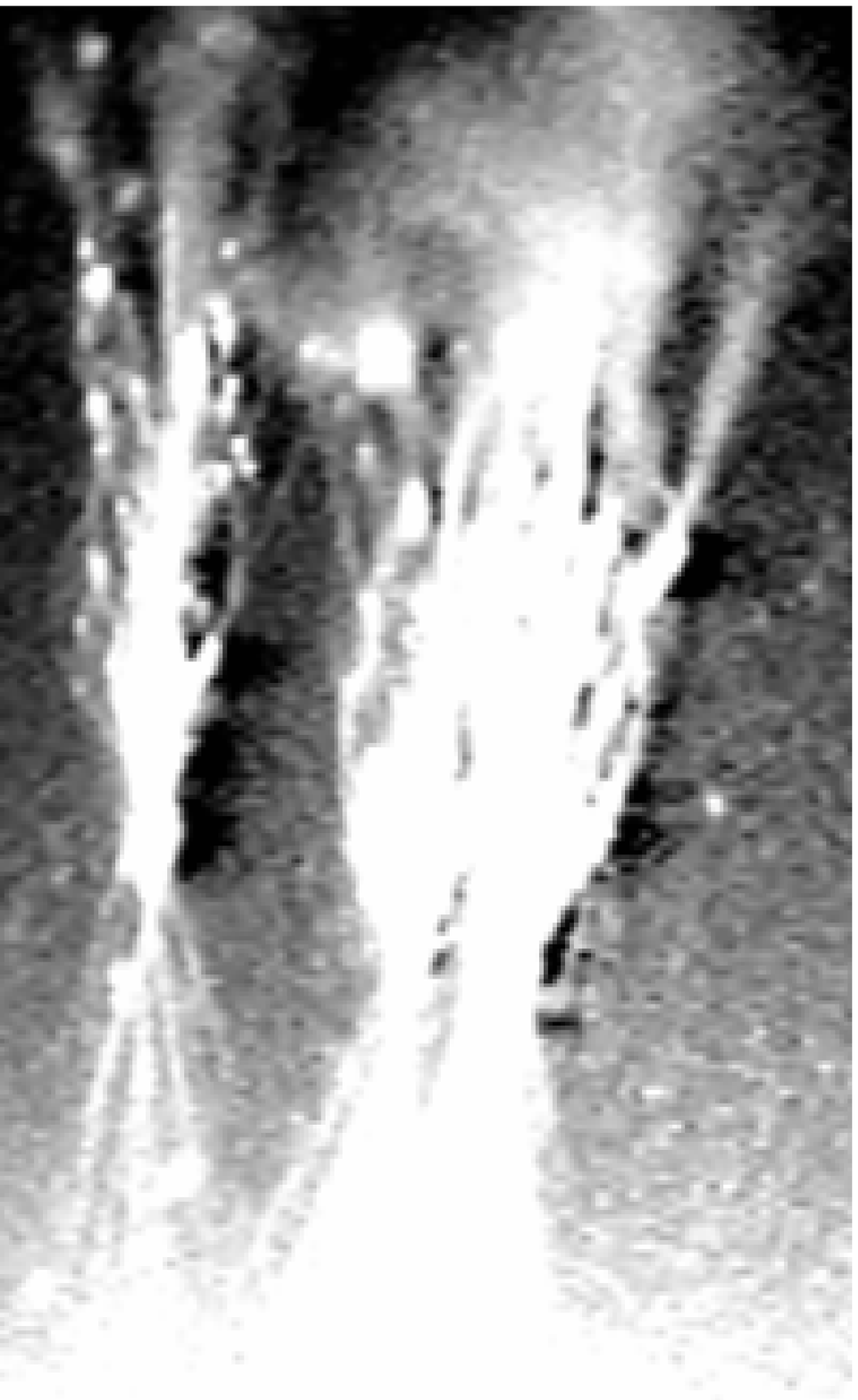}}}
    \put(120,185){\resizebox{100pt}{!}{
        \includegraphics*{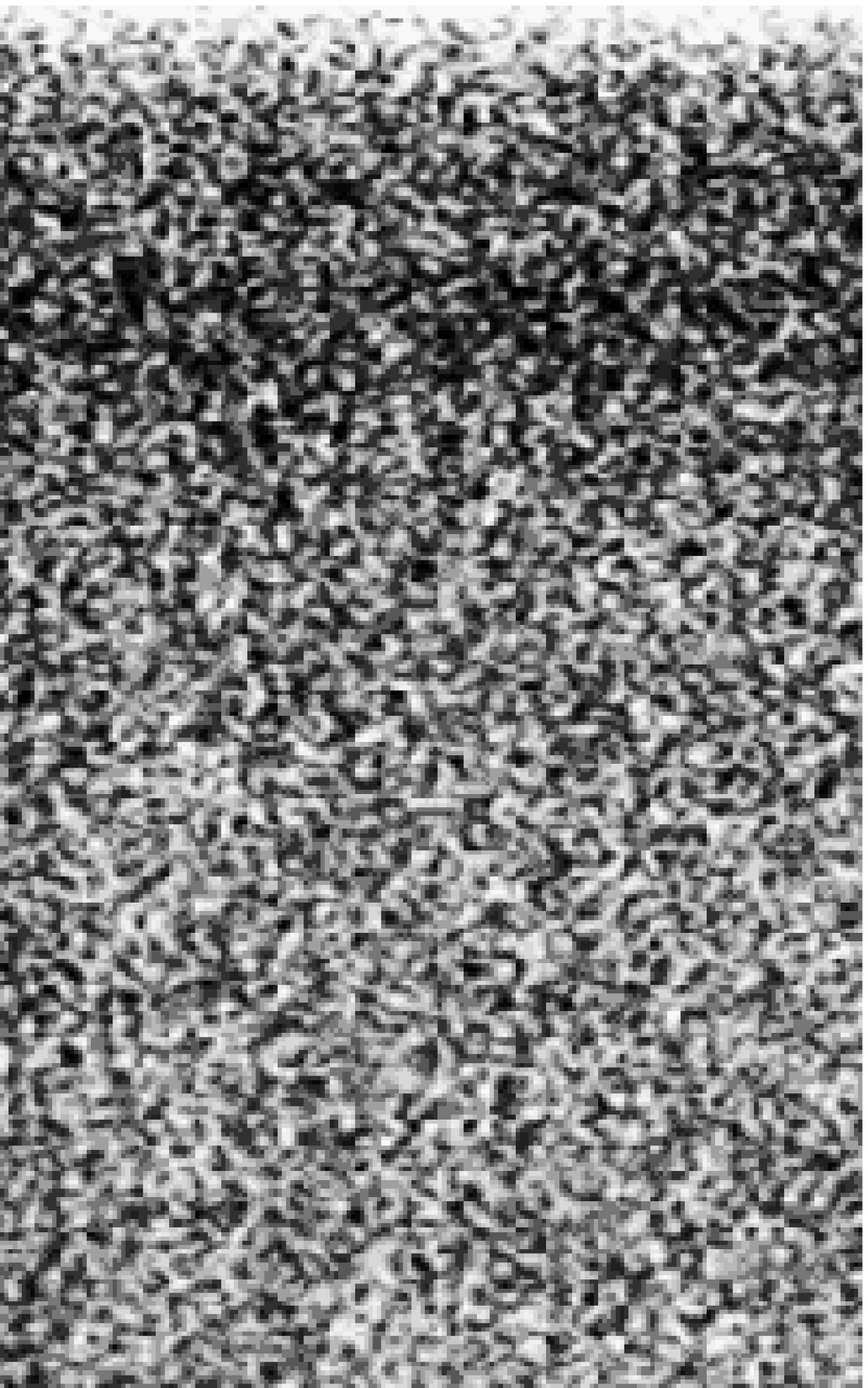}}}
    \thicklines
    \put(0,0){\dashline[+70]{3}(60,285)(100,285)}
    \put(0,0){\dashline[+70]{3}(60,283)(100,283)}
    \put(0,0){\dashline[+70]{3}(60,281)(100,281)}
    \put(0,0){\dashline[+70]{3}(60,284)(100,284)}
    \put(0,0){\dashline[+70]{3}(60,282)(100,282)}
    \put(0,350){a)}
    \put(0,160){c)}
    \put(120,350){b)}
  \end{picture}
  \caption{Blue Spectrum of July 28, 1998 06:41:01.278 Sprite.  (a) is
    the broadband image of the sprite with a black dashed line
    indicating the field of view of the slit spectrograph.  (b) is the
    histogram equalized 427.8~nm filtered imager data.  The imager
    detected no 427.8~nm emissions, agreeing with the lack of any such
    signature in the spectral observations.  (c) shows the spectrum
    observed by the NUV/Blue spectrograph as the solid line.  The
    dashed line is the synthetic fit to the observations including
    both neutral \nttp\ and ionized \nton.  The ionized contribution
    to the fit is shown as a dash-dot line (primarily near the 391.4~nm
    and 427.8~nm bands).}
  \label{fig:blue_spect}
\end{figure}

Sprites observed July 28, 1998 at 06:41:01.278 UT are shown in
Fig.~\ref{fig:blue_spect}.  The aircraft was flying at an altitude of
14,040~m at a great circle distance of 306~km from the positive
cloud-to-ground lightning discharge reported by the National Lightning
Detection Network temporally associated with the observed sprite.
Fig.~\ref{fig:blue_spect}~(a) is from the unfiltered imager (9.3$\deg$
x 7.0$\deg$ FOV).  The horizontal black band indicates the approximate
FOV of the slit in the NUV spectrograph.  After gamma correction and
background subtraction of the video image of the NUV spectrograph
output, three video fields were averaged together for a total time
integration of 40~ms (the spectral data were on the upper portion of
the scan read-out imager, the time integration only includes a portion
of the third field and is not the entire 50~ms of the three fields).
The 10 video scan lines over which spectral information was present
were also averaged, to improve the signal strength.
Fig.~\ref{fig:blue_spect}(c) presents these corrected observations
from the NUV spectrograph.  The observed spectrum is the solid line
while a synthetic spectral fit to the observations modeled with \nton\
and \nttp\ is shown as a dashed line in Fig.~\ref{fig:blue_spect}(c)
(the fit includes corrections for atmospheric transmission using MODTRAN \cite[]{morrill98} and
instrument response).  The dotted line in Fig.~\ref{fig:blue_spect}(c)
indicates the \nton\ contribution to the synthetic fit.  The synthetic
spectral code used to generate the fit are presented in detail by
\citet{bucsela97}.  Specifically, the vibrational distributions were
determined by free-fit of the band progressions - adjusted for
instrument response and atmospheric transmission - to the spectral
data.  An assumed rotational temperature of 230K was used, but given
the resolution, the fit is not very sensitive to temperature
\cite[]{bucsela99prep}.  Fig.~\ref{fig:blue_spect}(b) is the
observation of the sprite from the 427.8~nm filtered imager, which has
been histogram equalized to increase the contrast of the image.  No
indication of \nton \ 427.8~nm emissions from the sprite are apparent
(in agreement with the spectral fit shown in
Fig.~\ref{fig:blue_spect}(c)).



\section{Discussion}

Earlier spectral observations near the top of sprites, $\sim$~65~km, show
that the primary optical emission is the \ntop\ and here we confirm
that the blue component is predominantly \nttp\ .  Any weak molecular
ion emission will be short lived and only present in the streamer
heads know to makeup these events.  EXL98 studies
of filtered UV images indicate the \nttp\ / \nton\ 
ratio to be of order $4 \times 10^4$ in the 65~km region thus
confirming the lack of \nton\ emission \cite[]{morrill2002:e_energy}.  The current results and our
previous UV image ratio results indicate an estimated characteristic
energy of $\sim$~1.8~eV.  This reasonably matches the estimated energies
in the streamer body from the work of \citet{liu2005:LBH}.
Specifically, characteristic energy is the average energy of electrons
in a weakly ionized gas under an applied electric field.  If the gas
is collisional the energy distribution is non-Maxwellian.  The
characteristic energy is the $D_e/m_e$, where $D_e$ is the electron
transverse diffusion coefficient and $m_e$ is the mobility
\citep[cf.][p 21]{raizer91}.
These parameters are derived from laboratory by
observing the bulk diffusion and drift in an applied field.
Characteristic energies versus electric field for various gas are
tabulated in \cite{dutton75}.
Boltzmann solving models of sprites can
also report the average/characteristic energy as in 
\cite{morrill2002:e_energy}.
\cite{pasko97} did a cross-comparison between more complicated
Boltzmann-solving sprite models and those using empirical fits to
laboratory data and found generally good agreement between the two
methods.  These observations are highly time averaged relative to
natural sprite time scales and indicate that a significant portion of
the observed sprite emission is produced in the body of the streamers.
Unfortunately, the averaging of signal over multiple video frames was
necessary to achieve significant signal to noise for the synthetic
spectral fit.  It is important to recognize the different time scales
in sprites and related optical emissions--streamers with highly
ionized tips most likely create \nton\ emissions for sub-1~ms time
scales while a longer (10+~ms) \nttp\ emission is expected from sprite
bodies.

Other studies by \citet{Kuo:2005p2636} who studies \nttp\ / \nton\
ratios observed from space derived characteristic energies in the
range of 4.5-6.5eV below 60~km or 1.9 to 3.4 times the breakdown
field.  Our previous observations \citep{morrill2002:e_energy} yielded
energies at roughly the breakdown field so are significantly different
results which warrant further examination.  The 60~km altitude is
below our currently observed altitude.  Were observations made at this
altitude a significant \nton\ component would necessarily be present.
Additional blue/UV sprite spectral observations with wider altitude
range and higher temporal resolution are clearly needed.

\section{Conclusions}

We report observations and analysis of the blue/UV sprite spectrum at
65~km altitude.  The observed spectrum is well fitted by synthetic
spectral code (assuming a 230K rotational temperature) after
atmospheric transmission correction and system response correction.
The synthetic fit identifies only 2PG emission and implies electron
energies of roghly 1.8~eV.  This corresponds to similar results from
previous studies as well as generally agreeing with other model
results for electron energies in streamer bodies in sprites.  Because
of limitations in temporal resolution and vertical heigh of the
sprites in the rreported studie, additional blue/NUV observations are
needed to confirm these results.

\acknowledgements We thank the Remote Sensing Division of NRL for the
use of the UV intensified camera for the EXL98 flights.  Dan Osborne,
Jim Desroschers, Laura Peticolas, Veronika Besser, and Don Hampton
were instrumental to data collection and campaign operations.  Aeroair
Inc., and particularly Jeff Tobolsky, made all the EXL98 aircraft
missions fly.  The University of Alaska Fairbanks’ Geophysical
Institute group was supported by NASA Grants NAG5-5125 and NAG5-5019.
The work at NRL was sponsored by NASA NAG5-5172 and ONR.  Jeff
S. Morrill was partially supported by the Edison Memorial
graduate-training program at NRL.


\onecolumn
\begin{table}[t]  
\centerline{
\begin{tabular}{|c|cccccc|}
\hline
\multirow{2}{14mm}{Name} & Upper & Lower & 
\multirow{2}{14mm}{Lifetime} & 
Quench & Quench & Threshold \\
 & State & State & & Altitude & Particle & Energy \\
\hline
\ntop & \ntb & \nta & 6 $\mu$s & 53 km & \nt & 7.50 eV \\
\nttp & \ntc & \ntb & 50 ns & 30 km & \ot & 11.18 eV \\
\nton & \ntpb & \ntpx & 70 ns & 48 km & \nt, \ot & 18.56 eV \\
\ntm & \ntpa & \ntpx & 14 $\mu$s & 85-90 km & \nt & 16.54 eV \\
\ntvk & \nta & \ntx & 2 s & 145 km & \ox & 6.31 eV \\
\oton & \otpb & \otpx & 1.2 $\mu$s& 60 km & \nt & 18.2 eV \\
\hline
\end{tabular}}
\caption[Atmospheric Species]{Atmospheric Species.  Several neutral and ionized \nt \ and
\ot\ optical emissions that are observed in the aurora.  The earliest spectral
observations of sprites identified \ntop \ emissions, which are
the lowest threshold energy of the \nt \ states that emit optically.}
\label{tbl:atm_spec}
\end{table}
\twocolumn

\end{article}
\end{document}